\documentclass[twocolumn,aps,superscriptaddress,showpacs,floatfix,prc]{revtex4}
\usepackage{mathrsfs}
\usepackage{amssymb}
\usepackage{amsmath}
\usepackage{graphicx}
\usepackage[normalem]{ulem}
\usepackage[dvips]{color}
\usepackage{bm}
\usepackage{longtable}\usepackage{slashed}
\usepackage[utf8]{inputenc}
\usepackage{newtxtext}
\usepackage{enumitem}
\usepackage{empheq}
\renewcommand\theequation{\arabic{equation}}
\usepackage[varg]{txfonts}

\usepackage{epstopdf}

\usepackage{times}

\renewcommand\sout{\bgroup \color{red} \ULdepth=-.5ex \ULset}

\renewcommand{\rm}[1]{\textrm{#1}}
\renewcommand{\d}{\mathrm{d}}

\begin{document}

\title{Auxiliary Function Approach for Determining Symmetry Energy at Supra-saturation Densities}

\author{Bao-Jun Cai\footnote{bjcai87@gmail.com}}
\affiliation{Quantum Machine Learning Laboratory, Shadow Creator Inc., Shanghai 201208, China}
\author{Bao-An Li\footnote{Bao-An.Li$@$tamuc.edu}}
\affiliation{Department of Physics and Astronomy, Texas A$\&$M
University-Commerce, Commerce, TX 75429-3011, USA}
\date{\today}

\begin{abstract}
Nuclear symmetry energy $E_{\rm{sym}}(\rho)$ at density $\rho$ is normally expanded or simply parameterized as a function of $\chi=(\rho-\rho_0)/3\rho_0$ in the form of $E_{\rm{sym}}(\rho)\approx S+L\chi+2^{-1}K_{\rm{sym}}\chi^2+6^{-1}J_{\rm{sym}}\chi^3+\cdots$ using its magnitude $S$, slope $L $, curvature $K_{\rm{sym}}$ and skewness $J_{\rm{sym}}$ at the saturation density $\rho_0$ of nuclear matter. Much progress has been made in recent years in constraining especially the $S$ and $L$ parameters using various terrestrial experiments and astrophysical observations. However, such kind of expansions/parameterizations do not converge at supra-saturation densities where $\chi$ is not small enough, hindering an accurate determination of high-density $E_{\rm{sym}}(\rho)$ even if its characteristic parameters at $\rho_0$ are all well determined by experiments/observations. By expanding the $E_{\rm{sym}}(\rho)$ in terms of a properly chosen auxiliary function $\Pi_{\rm{sym}}(\chi,\Theta_{\rm{sym}})$ with a parameter $\Theta_{\rm{sym}}$ fixed accurately by an experimental $E_{\rm{sym}}(\rho_{\rm{r}})$ value at a reference density $\rho_{\rm{r}}$, we show that the shortcomings of the $\chi$-expansion can be completely removed or significantly reduced in determining the high-density behavior of $E_{\rm{sym}}(\rho)$. In particular, using two significantly different auxiliary functions, we show that the new approach effectively incorporates higher $\chi$-order contributions and converges to the same  $E_{\rm{sym}}(\rho)$ much faster than the conventional $\chi$-expansion at densities $\lesssim3\rho_0$. Moreover, the still poorly constrained skewness $J_{\rm{sym}}$ plays a small role in determining the $E_{\rm{sym}}(\rho)$ at these densities in the auxiliary function approach. The new approach thus provides a nearly model-independent constraint on the $E_{\rm{sym}}(\rho)$ at densities $\lesssim3\rho_0$. Several quantitative demonstrations using Monte Carlo simulations are given. 

\end{abstract}

\maketitle


\def\Ts{\Theta_{\rm{sym}}}
\def\Ps{\Pi_{\rm{sym}}(\chi,\Ts)}
\def\Es{E_{\rm{sym}}}
\def\Pso{\Pi_{\rm{sym}}(0,\Ts)}

\section{Introduction}\label{S1}

The density dependence of nuclear symmetry energy $E_{\rm{sym}}(\rho)$ is fundamental for addressing many important issues in both astrophysics and nuclear physics\,\cite{Lat00,ditoro,Steiner05,LCK08,Ditoro10,Chen11,Trau12,Tsang12,Tesym,Bal16,Oer17,BALI19}. While much progress has been made in constraining the $E_{\rm{sym}}(\rho)$ around the saturation density $\rho_0$ of  symmetric nuclear matter (SNM) over the last two decades, determining the $E_{\rm{sym}}(\rho)$ at supra-saturation densities remains a difficult problem.  Conventionally, the symmetry energy $E_{\rm{sym}}(\rho)$ predicted by nuclear many-body theories is often characterized by the first few coefficients of its Taylor expansion around $\rho_0$ in terms of $\chi=(\rho-\rho_0)/3\rho_0$, namely, 
\begin{equation}
E_{\rm{sym}}(\rho)\approx S+L\chi+\frac{1}{2}K_{\rm{sym}}\chi^2+\frac{1}{6}J_{\rm{sym}}\chi^3+\cdots
\label{EXPEsym}
\end{equation}
with the magnitude $S\equiv E_{\rm{sym}}(\rho_0)$, slope $L=[3\rho\d E_{\rm{sym}}/\d\rho]_{\rho_0}$,  curvature $K_{\rm{sym}}=[9\rho^2\d^2 E_{\rm{sym}}/\d\rho^2]_{\rho_0}$, and skewness $J_{\rm{sym}}=[27\rho^3\d^3 E_{\rm{sym}}/\d\rho^3]_{\rho_0}$.  While  in solving neutron star inverse-structure problems, the $E_{\rm{sym}}(\rho)$ function is not known {\it apriori}, it can be parameterized in the same form as above. The $S,L, K_{\rm{sym}}$ and 
$J_{\rm{sym}}$ are simply parameters to be determined from inverting data. Such kind of parameterizations are widely used in meta-modelings of nuclear equation of state (EOS), see, e.g., Refs.\,\cite{Steiner,Zhang18,MM1,Xie19,France1,Sof,Xie20,Con2,burg,Tsang,Bis21}. For example, these coefficients can be inferred from observational data through Bayesian statistical analyses or direct inversion techniques. Similar approaches have been used to constrain the SNM EOS using various observables from both terrestrial experiments and astrophysical observations. 

Despite of many fruitful applications of expansions and/or parametrizations using forms similar to the Eq.\,(\ref{EXPEsym}), the latter has one serious shortcoming. Namely, the dimensionless quantity $\chi$ becomes large as $\rho$ increases, the conventional expansion (\ref{EXPEsym}) breaks down eventually at high densities. Thus, it is inaccurate to predict the $E_{\rm{sym}}(\rho)$ at supra-saturation densities using Eq.\,(\ref{EXPEsym}) even if its first few characteristic parameters at $\rho_0$ are all well determined by experiments/observations. In this work, we explore possible ways to remedy this situation. In particular, using a properly chosen auxiliary function $\Pi_{\rm{sym}}(\chi,\Theta_{\rm{sym}})$ with a parameter $\Theta_{\rm{sym}}$ fixed accurately by an experimental $E_{\rm{sym}}(\rho_{\rm{r}})$ value at a reference density $\rho_{\rm{r}}$, one can expand the $E_{\rm{sym}}(\rho)$ as a function of $\widetilde{\nu}_{\rm{sym}}(\chi,\Ts)=\Pi_{\rm{sym}}(\chi,\Ts)-\Pi_{\rm{sym}}(0,\Ts)$. By performing Monte Carlo simulations with two auxiliary functions, we show that the auxiliary-function-based expansion can effectively incorporates higher $\chi$-order contributions and converges much faster than the conventional $\chi$-expansion at densities $\lesssim3\rho_0$, thus largely removing the shortcoming of the $\chi$-expansion in determining the high-density $E_{\rm{sym}}(\rho)$.

The rest of this paper is organized as follows. In the next section, we discuss in more detail the main problems of the $\chi$-expansion and our strategy to solve them for the purpose of determining $E_{\rm{sym}}(\rho)$ up to about $3\rho_0$ above which other degrees of freedom and/or phase transitions may have to be considered. In section \ref{S2}, the general framework and formalism of the auxiliary-function-based reconstruction of $E_{\rm{sym}}(\rho)$ are given. In section \ref{Test}, as a typical example, we show that the new approach at order $\widetilde{\nu}_{\rm{sym}}^3(\chi,\Ts)$ can successfully reconstruct the symmetry energy predicted by the relativistic mean-field (RMF) model with the FSUGold interaction up to about $5\rho_0$, while the conventional expansion already breaks down near $2\rho_0$. In section \ref{S3}, possible constraints on the high-density symmetry energy are given using two significantly different auxiliary functions by adopting three parameter sets different in their truncation orders and characteristics for the $E_{\rm{sym}}(\rho)$ at $\rho_0$. A brief summary and outlook are given in section \ref{S4}.

\section{The main issues and our strategies}\label{Issues}
During the last two decades, much progress has been made in constraining especially the low-order parameters of nuclear symmetry energy, e.g.,
the magnitude $S$ and the slope $L$ are relatively well constrained to the ranges of $S\approx31.7\pm3.2\,\rm{MeV}$ and $L\approx58.7\pm28.1\,\rm{MeV}$ \cite{Oer17,LiBA13} with few exceptions, respectively.
Moreover, the curvature $K_{\rm{sym}}$ is found to be effectively correlated with some neutron star properties\,\cite{Zhang18,Xie19,Xie20,YZhou19,YZhou19-a}, and its value has been shown to be negative. 
For instance, Bayesian analyses of the tidal deformation of canonical neutron stars from GW170817 and the radius data from NICER (Neutron Star Interior Explorer) found a value of $K_{\mathrm{sym}}=-120_{-100}^{+80}$ at 68\% confidence level\,\cite{Xie20}. In addition,  a recent Bayesian analysis on some theoretical calculations gave a skewness $J_{\rm{sym}}\approx90\pm 334$\,MeV at 68\% confidence level\,\cite{Som20}, see also the constraint $J_{\rm{sym}}\approx296.8\pm73.6\,\rm{MeV}$ by analyzing the systematics of over 520 energy density functionals\,\cite{Mon17}, indicating that the skewness $J_{\rm{sym}}$ is probably positive.

Given the above information about the characteristics of $E_{\rm{sym}}(\rho)$ at $\rho_0$, the expression (\ref{EXPEsym}) provides the simplest way to predict the symmetry energy at supra-saturation densities. However, to what accuracies and up to what densities it can be used have been uncertain because the dimensionless quantity $\chi$ becomes eventually large and then the $\chi$-expansion diverges as $\rho$ increases. Thus, despite of the progress made in constraining the characteristics $S, L, K_{\rm{sym}}$ and even the $J_{\rm{sym}}$, there still remain some fundamental issues related to the expansion (\ref{EXPEsym}). Moreover, a few natural questions concerning the structure and implications of the expansion (\ref{EXPEsym}) emerge:
\begin{enumerate}[leftmargin=*,label=(\alph*)]
\item Are the characteristics $(S,L,K_{\rm{sym}},J_{\rm{sym}},\cdots)$ enough
to describe the symmetry energy at supra-saturation densities, such as $\rho\approx2\rho_0$ or $\rho\approx3\rho_0$?
Are some of them irrelevant for $E_{\rm{sym}}(\rho)$ at densities up to $3\rho_0$?  The $E_{\rm{sym}}(\rho)$ around $(1-3)\rho_0$ is most important for the radii and tidal deformations of canonical neutron stars but is currently poorly determined\,\cite{BALI19}. Above this density range, non-nucleonic degrees of freedom become important and various phase transitions may set in.
The answer to the last question is definitely ``no'' in the conventional expansion (\ref{EXPEsym}) since $\chi=2/3$ at $\rho=3\rho_0$ and a small change of the skewness $J_{\rm{sym}}$ may easily introduce sizable effects on $E_{\rm{sym}}(3\rho_0)$.
\item  Can we find other forms to re-express the symmetry energy to make the corresponding expansion
based on the same quantities $(S,L,K_{\rm{sym}},J_{\rm{sym}},\cdots)$ quickly converge and more accurately describe the symmetry energy at supra-saturation densities? It does not mean that exact functionals and/or theories for the high-density asymmetric nuclear matter (ANM) EOS are not needed.  Our main purpose is to find once the characteristics like $L$ and $K_{\rm{sym}}$ are well constrained, if we can make some (near) model-independent predictions for the $E_{\rm{sym}}(\rho)$ at densities $\lesssim2\rho_0$ or $3\rho_0$? Intuitively, it should be, however it is known that different models often predict (very) different high-density behaviors for $E_{\rm{sym}}(\rho)$ even when they predict very similar or the same characteristics for $E_{\rm{sym}}(\rho)$ at $\rho_0$. In this sense, we would like to study if we can ``reconstruct'' accurately the symmetry energy at supra-saturation densities based on its known characteristics at $\rho_0$ constrained by experiments/observations instead of calculating it based on any nuclear many-body theory.
\item In searching for new forms of $f(\rho)$ to reconstruct the symmetry energy $E_{\rm{sym}}(\rho)=S+f(\rho)$, a natural boundary condition is that the first several terms of $f(\rho)$ expanded around $\chi=0$ should be the same as the ones given by the conventional expansion (\ref{EXPEsym}). However, to go beyond the latter, certain higher $\chi$-order contributions should also be effectively encapsulated in $f(\rho)$ using still only the first few characteristics of $E_{\rm{sym}}(\rho)$ at $\rho_0$, i.e., $L, K_{\rm{sym}}$ and $J_{\rm{sym}}$.
\end{enumerate}

The breakdown of the conventional expansions (\ref{EXPEsym}) is due to the fact that the $\chi$ is not always small enough for small-quantity expansions. 
Even if one re-scales the $\chi$ to be $\widetilde{\chi}=(\rho-\rho_0)/\xi\rho_0$ with $\xi\geq3$ a constant, the corresponding terms in (\ref{EXPEsym}) still have their original forms, e.g., $2^{-1}K_{\rm{sym}}\chi^2\to(\xi^2/18)K_{\rm{sym}}\widetilde{\chi}^2$. The adjusted expansion shares the same shortcomings as the original one.
On the other hand, if we adopt an effective auxiliary function of $\chi$, i.e., $
\Pi_{\rm{sym}}(\chi,\Theta_{\rm{sym}})$ with $\Theta_{\rm{sym}}$ a model parameter to be determined at a reference density $\rho_{\rm{r}}$ where the $E_{\rm{sym}}(\rho_{\rm{r}})$ is well constrained by experimental data, 
and then expand the symmetry energy around the difference $\widetilde{\nu}_{\rm{sym}}(\chi,\Ts)=\Pi_{\rm{sym}}(\chi,\Ts)-\Pi_{\rm{sym}}(0,\Ts)$,
\begin{equation}
\frac{\d^nE_{\rm{sym}}}{\d\rho^n}(\rho-\rho_0)^n\rightarrow
\frac{\d^nE_{\rm{sym}}}{\d\Pi_{\rm{sym}}^n}\widetilde{\nu}_{\rm{sym}}^n(\chi,\Ts),\end{equation}
some new possibilities emerge:
\begin{enumerate}[leftmargin=*,label=(\alph*)]
\item If the function $\Pi_{\rm{sym}}(\chi,\Theta_{\rm{sym}})$ is selected well, then expanding it around $\chi\approx0$ gives,
\begin{align}
\Pi_{\rm{sym}}(\chi,\Theta_{\rm{sym}})\approx& \Pi_{\rm{sym}}(0,\Theta_{\rm{sym}})+\Pi'_{\rm{sym}}(0,\Theta_{\rm{sym}})\chi\notag\\
&+\Pi''_{\rm{sym}}(0,\Theta_{\rm{sym}})\chi^2/2\notag\\
&+\Pi'''_{\rm{sym}}(0,\Theta_{\rm{sym}})\chi^3/6+\cdots,\end{align}
where the prime ``$'$'' denotes derivatives with respect to $\chi$ (or equivalently with respect to the density $\rho$).
Although the symmetry energy is expanded, e.g., to order $\chi^3$,
the $\Pi_{\rm{sym}}(\chi,\Theta_{\rm{sym}})$ can effectively generates higher order terms in $\chi$.
If the factor $\widetilde{\nu}_{\rm{sym}}(\chi,\Ts)$ is small enough at supra-saturation densities, the auxiliary function expansion is naturally expected to
converge faster than the conventional expansion (\ref{EXPEsym}). 
Of course, a reasonable auxiliary function $\Pi_{\rm{sym}}(\chi,\Theta_{\rm{sym}})$ has to be chosen to meet this goal.

\item The question of model dependence related to choosing the auxiliary function $\Pi_{\rm{sym}}(\chi,\Theta_{\rm{sym}})$ and determining the associated $\Theta_{\rm{sym}}$ parameter emerges naturally since the auxiliary function could take vastly different forms. Thus, one should compare results of using very different auxiliary functions. Once a model $\Pi_{\rm{sym}}(\chi,\Theta_{\rm{sym}})$ is adopted/selected, the parameter $\Theta_{\rm{sym}}$ can be determined by the symmetry energy at a density where it is well determined, similar to determining the low-energy coefficients in chiral effective field theories by some low-energy scattering processes\,\cite{Bur21}. Nevertheless, logically one should self-consistently determine the values of $S, L, K_{\rm{sym}},\cdots$ and $\Ts$ using a certain selected auxiliary function model simultaneously (via analyzing nuclear experimental data and/or astrophysical observations) and see how the symmetry energy depends on the form of the auxiliary function (at supra-saturation densities). As the first step in our exploratory study in this direction, here we merely investigate whether one could reconstruct the $E_{\rm{sym}}(\rho)$ at supra-saturation densities $\lesssim3\rho_0$ in an effective manner once the lower-order characteristics are known without caring about how these characteristics are constrained. Future works along this line should consider more about the self-consistency of the approach. 
\end{enumerate}

In the following, we use two significantly different auxiliary functions, namely an exponential and an algebraic model. Interestingly, we find that the predicted symmetry energies at supra-saturation densities $\rho_0\lesssim\rho\lesssim3\rho_0$ are almost the same with the two models, indicating that the auxiliary-function-based reconstruction is effective and has some essential universality. Moreover, in our Monte Carlo simulations the auxiliary-function-based expansion indeed converges faster than the conventional $\chi$-expansion. 

\section{Framework and Formalism}\label{S2}
Given the four characteristic parameters $S, L, K_{\rm{sym}}$ and $J_{\rm{sym}}$ of $E_{\rm{sym}}(\rho)$ at $\rho_0$,  the symmetry energy $E_{\rm{sym}}(\rho)$ can be expanded around $\Pi_{\rm{sym}}(\chi,\Theta_{\rm{sym}})=\Pi_{\rm{sym}}(0,\Theta_{\rm{sym}})$ to order $\nu_{\rm{sym}}^3(\chi,\Ts)$ as
\begin{align}
E_{\rm{sym}}(\rho)\approx&S+L\nu_{\rm{sym}}(\chi,\Theta_{\rm{sym}})
+\frac{1}{2}K_{\rm{sym}}\Phi\nu^2_{\rm{sym}}(\chi,\Theta_{\rm{sym}})\notag\\
&+\frac{1}{6}J_{\rm{sym}}\Psi\nu^3_{\rm{sym}}(\chi,\Theta_{\rm{sym}}),\label{EXPEsymNEW}
\end{align}
where 
\begin{align}
\Phi=&1
+\left.\frac{L}{K_{\rm{sym}}}\left({\displaystyle\frac{1}{3\rho}\frac{\partial^2\rho}{\partial\Pi_{\rm{sym}}^2}}\right)\right/{\displaystyle\left(\frac{1}{3\rho}\frac{\partial\rho}{\partial\Pi_{\rm{sym}}}\right)^2}_{\chi=0},\label{def_PHI}\\
\Psi=&1+\left.\frac{K_{\rm{sym}}}{J_{\rm{sym}}}
\left({\displaystyle\frac{1}{3\rho^2}\frac{\partial\rho}{\partial\Pi_{\rm{sym}}}\frac{\partial^2\rho}{\partial\Pi_{\rm{sym}}^2}}\right)\right/{\displaystyle\left(\frac{1}{3\rho}\frac{\partial\rho}{\partial\Pi_{\rm{sym}}}\right)^3}_{\chi=0}\notag\\
&\left.+\frac{L}{J_{\rm{sym}}}\left({\displaystyle\frac{1}{3\rho}\frac{\partial^3\rho}{\partial\Pi_{\rm{sym}}^3}}\right)\right/{\displaystyle\left(\frac{1}{3\rho}\frac{\partial\rho}{\partial\Pi_{\rm{sym}}}\right)^3}_{\chi=0},\label{def_PSI}
\end{align}
and,
\begin{equation}\label{def_nusymkk}
\nu_{\rm{sym}}(\chi,\Theta_{\rm{sym}})\equiv \left[\frac{1}{3\rho}\frac{\partial\rho}{\partial\Pi_{\rm{sym}}(\chi,\Theta_{\rm{sym}})}\right]_{\chi=0}\cdot\widetilde{\nu}_{\rm{sym}}(\chi,\Theta_{\rm{sym}}),
\end{equation}
where $\widetilde{\nu}_{\rm{sym}}(\chi,\Ts)=\Pi_{\rm{sym}}(\chi,\Ts)-\Pi_{\rm{sym}}(0,\Ts)$.
It can be proved straightforwardly that the conventional expansion (\ref{EXPEsym}) corresponds to the special case of selecting $\Pi_{\rm{sym}}(\chi,\Ts)\propto \chi$. 
However, terms higher than $\chi^3$ are effectively included in (\ref{EXPEsymNEW}) even it is truncated at order $\nu_{\rm{sym}}^3(\chi,\Ts)$, since the latter itself encapsulates the higher order effects in $\chi$.
Moreover, the $K_{\rm{sym}}\Phi$ and $J_{\rm{sym}}\Psi$ in (\ref{EXPEsymNEW}) can be treated as the effective curvature and skewness of the symmetry energy with respect to the expansion in $\nu_{\rm{sym}}(\chi,\Ts)$.

While the expansion (\ref{EXPEsymNEW}) is general, its applications depend on the specific form of $\Pi_{\rm{sym}}(\chi,\Theta_{\rm{sym}})$ to be adopted.
In the following, we consider two models, namely the exponential model (abbreviated as ``exp'') and the algebraic model (abbreviated as ``alge''). Specifically,
\begin{enumerate}[leftmargin=*]
\item In the exponential model, $\Pi_{\rm{sym}}(\chi,\Theta_{\rm{sym}})=\exp[-\Theta_{\rm{sym}}(1+3\chi)]$, the $\nu_{\rm{sym}}$ is given as
\begin{equation}
\nu_{\rm{sym}}(\chi,\Theta_{\rm{sym}})=\frac{1}{3\Theta_{\rm{sym}}}
\left(1-e^{-3\chi\Theta_{\rm{sym}}}\right).\label{nnn-k}
\end{equation}
The resulting auxiliary-function-based reconstruction of the symmetry energy is given by
\begin{align}
E_{\rm{sym}}(\rho)\approx&S+L\nu_{\rm{sym}}(\chi,\Theta_{\rm{sym}})\notag\\
&+\frac{1}{2}K_{\rm{sym}}\left(1+\frac{3L}{K_{\rm{sym}}}\Ts\right)\nu_{\rm{sym}}^2(\chi,\Theta_{\rm{sym}})\notag\\
&+\frac{1}{6}J_{\rm{sym}}\left(1+\frac{9K_{\rm{sym}}}{J_{\rm{sym}}}\Ts
+\frac{18L}{J_{\rm{sym}}}\Ts^2
\right)\notag\\
&\hspace*{1.cm}\times\nu_{\rm{sym}}^3(\chi,\Theta_{\rm{sym}})\label{tt1}.
\end{align}
Some new features emerge in (\ref{tt1}). Firstly, besides the conventional term $2^{-1}K_{\rm{sym}}
$, a new term $3\Ts L/K_{\rm{sym}}$ (normalized by $2^{-1}K_{\rm{sym}}$) contributes at order $\nu_{\rm{sym}}^2(\chi,\Ts)$. This term is generally sizable and can not be thought as a perturbation.
Secondly, for small $\chi$, e.g, $\rho_0\lesssim\rho\lesssim3\rho_0$, we have
\begin{align}\label{dk}
\nu_{\rm{sym}}(\chi,\Ts)\approx&\chi-\frac{3}{2}\Ts\chi^2+\frac{3}{2}\Ts^2\chi^3-\frac{9}{8}\Ts^3\chi^4\notag\\
&+\frac{27}{40}\Ts^4\chi^5-\frac{27}{80}\Ts^5\chi^6+\cdots,\notag\\
\to&\chi,~~\chi\to0,
\end{align}
i.e., although high-order terms such as the fourth-order kurtosis $I_{\rm{sym}}$, etc., are absence in the expansion (\ref{EXPEsymNEW}),
the effects of $\chi^4$ or $\chi^5$ are effectively generated.
It means that the effects of high order terms are modeled with the help
of the function $\Pi_{\rm{sym}}(\chi,
\Theta_{\rm{sym}})$.
For example,  the effective kurtosis $I_{\rm{sym}}^{\rm{eff}}$ of the symmetry energy defined as the fourth-order Taylor's expansion coefficient of $E_{\rm{sym}}(\rho)$ at $\rho_0$ could be obtained in terms of $L, K_{\rm{sym}}, J_{\rm{sym}}$ and $\Ts$ once the expression (\ref{tt1}) is expand to order  $\chi^4$,  i.e., $
I_{\rm{sym}}^{\rm{eff}}=-9\Ts(2J_{\rm{sym}}+11\Ts K_{\rm{sym}}+18\Ts^2L)$.  From the above expansion, one can see that the limit $\Ts\to0$ (equivalently $\Pi_{\rm{sym}}(\chi,
\Theta_{\rm{sym}})\to\chi$)  is equivalent to the conventional expansion of the $E_{\rm{sym}}(\rho)$ in Eq. (\ref{EXPEsym}).
\item In the algebraic model, $\Pi_{\rm{sym}}(\chi,\Theta_{\rm{sym}})=[1+\Theta_{\rm{sym}}(1+3\chi)]^{-1}$, we have
\begin{equation}
\nu_{\rm{sym}}(\chi,\Theta_{\rm{sym}})=\chi\frac{1+\Theta^{-1}_{\rm{sym}}}{1+3\chi+\Theta^{-1}_{\rm{sym}}},\label{nnn-1}
\end{equation}
for the expansion element, and
\begin{align}
E_{\rm{sym}}(\rho)\approx&S+L\nu_{\rm{sym}}(\chi,\Theta_{\rm{sym}})\notag\\
&+\frac{1}{2}K_{\rm{sym}}\left(1+\frac{6L}{K_{\rm{sym}}}\frac{1}{1+\Theta^{-1}_{\rm{sym}}}\right)\nu_{\rm{sym}}^2(\chi,\Theta_{\rm{sym}})\notag\\
&+\frac{1}{6}J_{\rm{sym}}\Bigg[1+\frac{18K_{\rm{sym}}}{J_{\rm{sym}}}\frac{1}{1+\Theta^{-1}_{\rm{sym}}}\notag\\
&\hspace*{0.5cm}+\frac{54L}{J_{\rm{sym}}}\left(\frac{1}{1+\Theta^{-1}_{\rm{sym}}}\right)^2
\Bigg]\nu_{\rm{sym}}^3(\chi,\Theta_{\rm{sym}})\label{tt2}
\end{align}
for the auxiliary-function-based reconstruction of the symmetry energy.
There are two limits for the function $\nu_{\rm{sym}}$.
If $\Ts$ is small (i.e., the small-$\Ts$ limit), we then have
\begin{equation}
\nu_{\rm{sym}}(\chi,\Ts)\approx\chi\left(1-3\chi\Ts\right)\to\chi,\end{equation}
where the last relation holds for $\chi\to0$ (near the saturation density).
On the other hand, if $\Ts$ is large (i.e., the large-$\Ts$ limit), then one can treat $1/\Ts$ as a small quantity,
\begin{equation}
\nu_{\rm{sym}}(\chi,\Ts)\approx\frac{\chi}{1+3\chi}\left(1+\frac{1}{\Ts}\frac{3\chi}{1+3\chi}\right)\to\chi.
\end{equation}
In fact from expressions (\ref{nnn-1}) and (\ref{tt2}), one can find that the $\Ts$ appears naturally in the form of $1/\Ts$.
\end{enumerate}

There are two different approaches to determine the parameter $\Ts$. In the first approach, one determines the expression for $\Ts$ by truncating the symmetry energy in the expansion (\ref{EXPEsymNEW}). For example, if the symmetry energy is truncated at order $\nu_{\rm{sym}}$,  i.e., $E_{\rm{sym}}(\rho)\approx S+L\nu_{\rm{sym}}(\chi,\Ts)$, then by using the curvature $K_{\rm{sym}}$ (which is assumed to be known) one can obtain the expression for $\Ts$.
For instance,  the $\Ts=-K_{\rm{sym}}/3L$ in the exponential model is obtained according to the expression (\ref{dk}), i.e., $E_{\rm{sym}}(\rho)\approx S+L\chi-3L\Ts\chi^2/2+\mathcal{O}(\chi^3)$.
This determination could also be expressed as the following condition,
\begin{equation}\label{ur-2}
\Phi=0,
\end{equation}
with $\Phi$ defined in Eq. (\ref{def_PHI}), see the appendix for a general proof of this equivalence. In the exponential model, the expression for $\Phi$ is given by $\Phi=1+3\Ts L/K_{\rm{sym}}$, and setting it to zero gives the $\Ts=-K_{\rm{sym}}/3L$.
At this order, we thus have $E_{\rm{sym}}(\rho)\approx S-L^2K^{-1}_{\rm{sym}}[1-e^{\chi K_{\rm{sym}}/L}]$.
Similarly, the $\Ts$ could be determined when the truncation of the symmetry energy (\ref{EXPEsymNEW}) is made at order $\nu_{\rm{sym}}^2$, by solving the equation for $\Ts$ from
\begin{equation}
\Psi=0
\end{equation}
with $\Psi$ defined in Eq. (\ref{def_PSI}).
For the exponential model,  for example, we have $\Ts=-[K_{\rm{sym}}/4L][1\pm(1-8LJ_{\rm{sym}}/9K_{\rm{sym}}^2)^{1/2}]$ under the condition $J_{\rm{sym}}\leq9K_{\rm{sym}}^2/8L$, since now the symmetry energy (\ref{EXPEsymNEW}) is expanded as $E_{\rm{sym}}(\rho)\approx S+L\chi+2^{-1}K_{\rm{sym}}\chi^2-3\Ts\chi^3(2L\Ts+K_{\rm{sym}})/2$.
Setting the last term equal to $6^{-1}J_{\rm{sym}}\chi^3$ gives then the expression for $\Ts$ at this order.

The second approach which is adopted in the current work determines the value of $\Ts$ by some empirical value of the symmetry energy at a density where there are sufficiently accurate experimental and/or theoretical constraints. Here we use\,\cite{JXu20}
\begin{equation}\label{fit-sk}
E_{\rm{sym}}(\rho_{\rm{low}})\approx16.4\pm0.5\,\rm{MeV},~~\rho_{\rm{low}}\approx0.05\,\rm{fm}^{-3}.
\end{equation}
This empirical value was recently extracted from Bayesian analyses of both the centroid energy of isovector giant dipole resonance and electrical dipole polarizability of $^{208}$Pb\,\cite{JXu20}. Of course, other empirical constraints on the symmetry energy mostly below $\rho_0$ could also be used for determining the $\Ts$. 
For example, the $E_{\rm{sym}}(\rho_{\rm{c}})=26.65\pm0.20\,\rm{MeV}$\,\cite{Zha13} at the so-called cross density $\rho_{\rm{c}}\approx0.11\,\rm{fm}^{-3}$ where many different model predictions for $E_{\rm{sym}}(\rho)$ using various effective interactions cross is another useful point. 

The physical meaning and determination procedure of $\Ts$ discussed above can be seen more clearly by making an analogy with solving the forced oscillator problem. 
Consider an oscillator moving under an extra force $-\sigma x^3$ besides the conventional Hooke's force $f_{\rm{H}}(x)=-kx$ with $k$ the spring constant, i.e., $f_{\rm{tot}}(x)=-kx-\sigma x^3$, here $\sigma>0$ is a high-order coefficient. In order to avoid dealing with the dynamical variable ``$x$'' in the nonlinear term $\sigma x^3$, one can define an effective spring constant $k_{\rm{eff}}\approx k(1+s_1\phi+s_2\phi^2)+\mathcal{O}(\phi^3)$ where $\phi=\sigma d_{\max}^2/k\ll1$ with $d_{\max}$ the amplitude of the oscillator, via the equation of energy conservation $2^{-1}m\dot{x}^2+2^{-1}kx^2+U_{\sigma}(x)=2^{-1}kd_{\max}^2+U_{\sigma}(d_{\max}),U_{\sigma}(x)=4^{-1}\sigma x^4$. The two coefficients $s_1$ and $s_2$ could be matched by requiring, e.g., that the periods of the oscillation obtained by using the full potential $U_{\rm{tot}}(x)=2^{-1}kx^2+U_{\sigma}(x)$ and the effective potential $U_{\rm{eff}}(x)=2^{-1}k_{\rm{eff}}x^2$ are the same, to order $\phi^2$.
Consequently, one obtains $s_1=3/4$ and $s_2=-3/128$ and thus $k_{\rm{eff}}\approx k(1+3\phi/4-3\phi^2/128)$ through the basic formula $T=2\pi(m/k_{\rm{eff}})^{1/2}$, as the period of the oscillation from the full potential is given by
\begin{equation}
T\approx2\pi\sqrt{\frac{m}{k}}\times\left(1-\frac{3}{8}\frac{\sigma d_{\max}^2}{k}+\frac{57}{256}\frac{\sigma^2d_{\max}^4}{k^2}\right)
\end{equation}
to order $\phi^2$.
In other applications, one can then use the effective potential $U_{\rm{eff}}(x)=2^{-1}k_{\rm{eff}}x^2$ to do the relevant calculations (without dealing with the dynamical variable ``$x$'').
The high order effects characterized
by the parameter $\sigma$ appear in the effective potential through the low-order coefficient $k_{\rm{eff}}$.
The effective spring constant $k_{\rm{eff}}$ could be constructed order by order with respect to the perturbative element $\phi$, similar to the construction of the $\Ts$ parameter by considering certain types of higher order contributions from the $K_{\rm{sym}},J_{\rm{sym}}$, etc.

\section{Testing the auxiliary-function-based approach against a known $E_{\textmd{sym}}(\rho)$ Functional}\label{Test}

 \renewcommand*\figurename{\small Fig.}
\begin{figure}[h!]
\centering
\includegraphics[width=6.5cm]{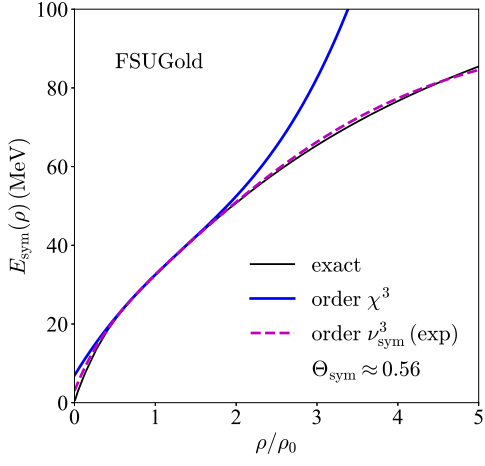}
 \caption{(Color Online). Nuclear symmetry energy predicted by the nonlinear relativistic mean field model with the FSUGold interaction in comparison with its reconstructions from the conventional and auxiliary-function-based 
 expansions.}\label{fig_EsymFSUre}
\end{figure}

\renewcommand*\figurename{\small Fig.}
\begin{figure*}
\centering
\includegraphics[width=5.2cm]{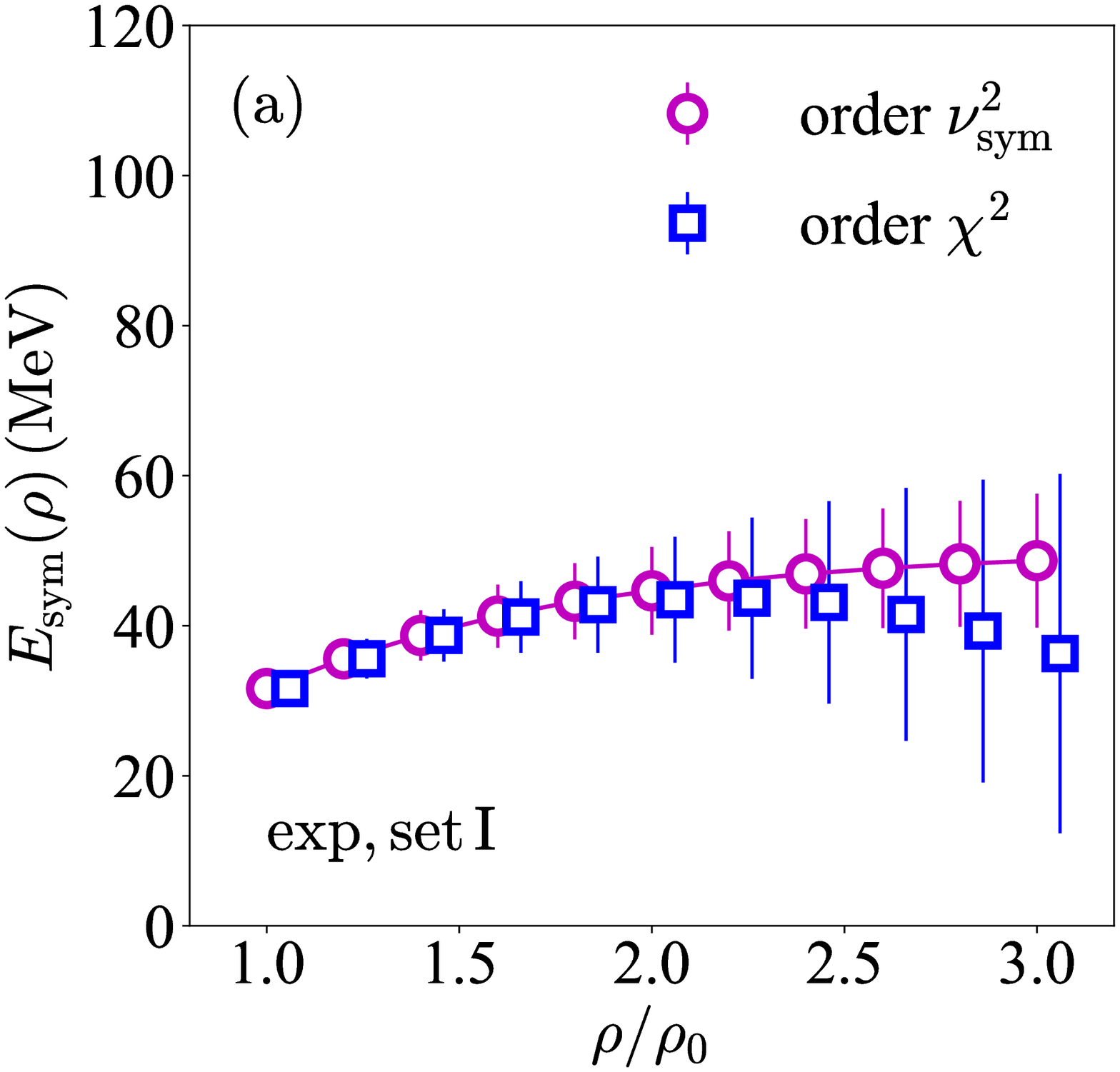}\qquad
\includegraphics[width=5.2cm]{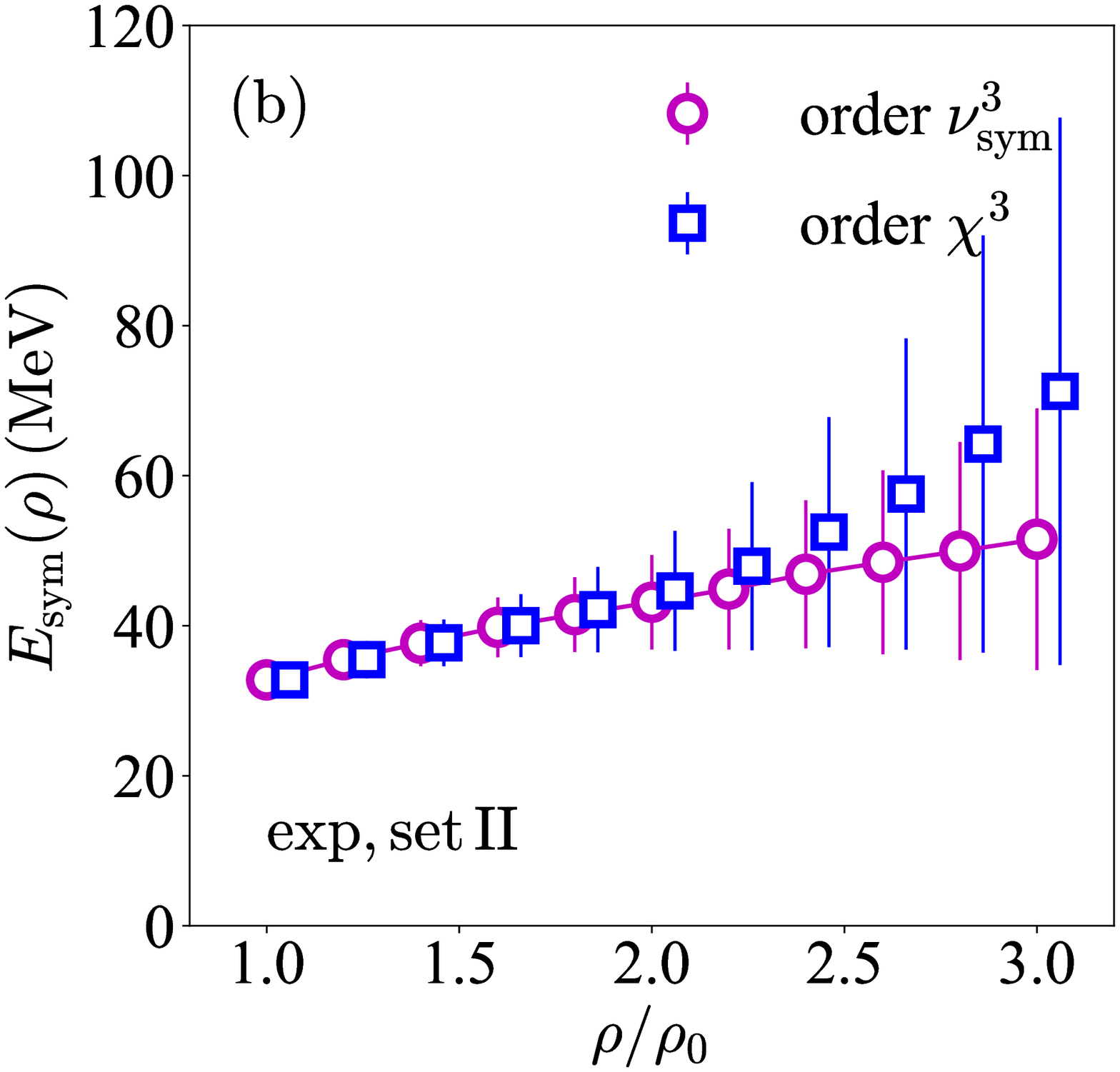}\qquad
\includegraphics[width=5.2cm]{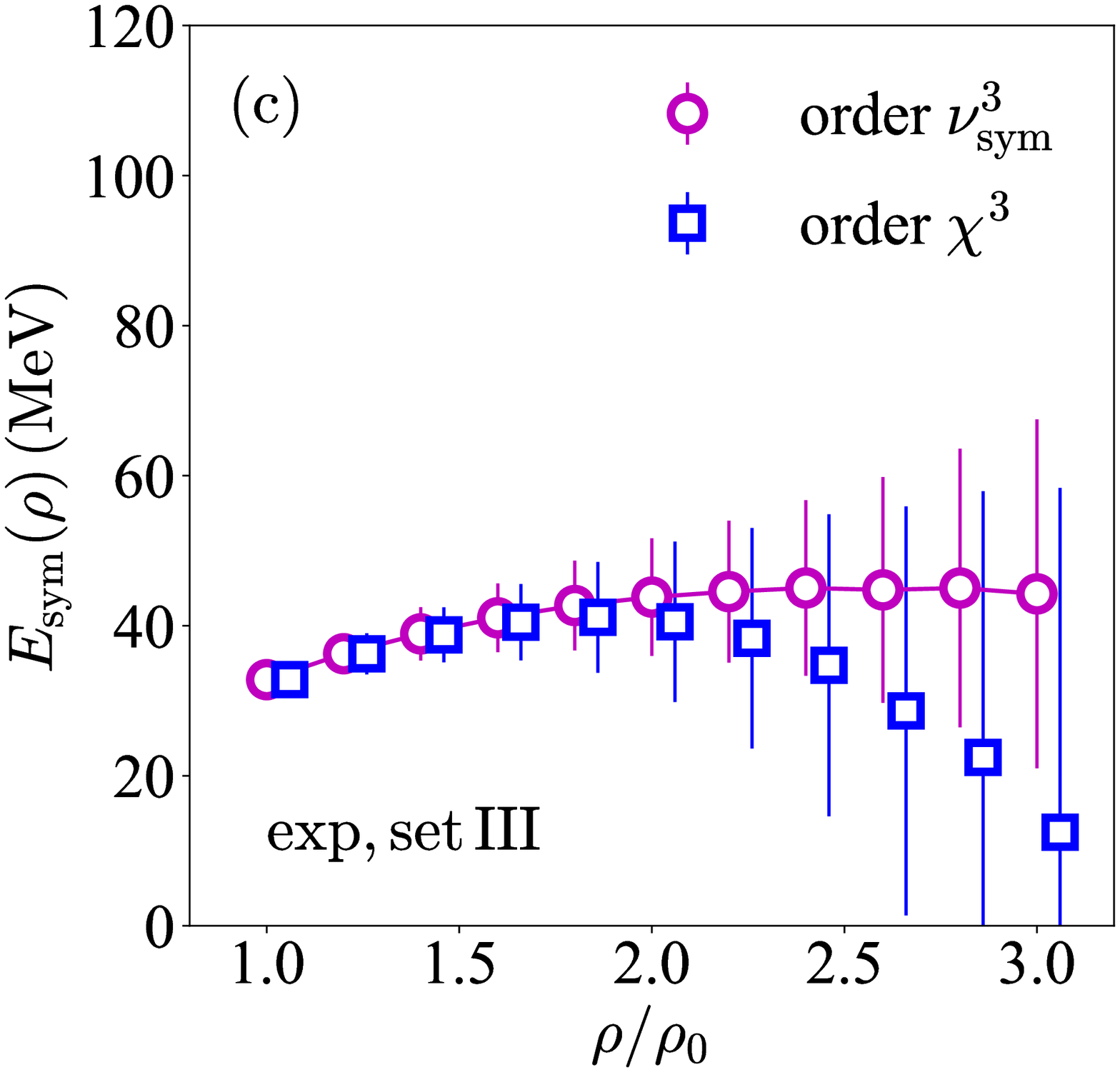}
 \caption{(Color Online). Density dependence of nuclear symmetry energy from simulations adopting the exponential model (abbreviated as ``exp'' in the figure) for the auxiliary function for the test set I (panel (a)), II (panel (b)), and III (panel (c)), respectively.}\label{fig_Esymu-set-ABC}
\end{figure*}

Again, our main goal is to see if and how one can reconstruct accurately the high-density behavior of nuclear symmetry by using its characteristics at saturation density. 
As the first test of the auxiliary-function-based reconstruction, we show in Fig.\,\ref{fig_EsymFSUre} the $E_{\rm{sym}}(\rho)$ obtained
by the expansion (\ref{tt1}) within the exponential model to order $\nu_{\rm{sym}}^3$, and the conventional expansion to order $\chi^3$, with respect to the RMF prediction using the FSUGold interaction\,\cite{Tod05}. With the FSUGold parameters, the RMF predicts a saturation density $\rho_0\approx0.148\,\rm{fm}^{-3}$ for SNM as well as a magnitude $S\approx32.5\,\rm{MeV}$, slope $L\approx60.4\,\rm{MeV}$, curvature $K_{\rm{sym}}\approx-51.0\,\rm{MeV}$ and a skewness $J_{\rm{sym}}\approx426.5\,\rm{MeV}$ for the symmetry energy. The $\Ts$ parameter in this example is found to be about 0.56 within the exponential model, and consequently, the correction $3L\Ts/K_{\rm{sym}}$ in (\ref{tt1}) gives a value about $-2$, which is obviously non-perturbative.
Similarly, the correction $9K_{\rm{sym}}\Ts/J_{\rm{sym}}+18L\Ts^2/J_{\rm{sym}}$ in (\ref{tt1}) generates a value of about 0.2. They both are thus important in reconstructing the $E_{\rm{sym}}(\rho)$ in a broad density range.

Interestingly, it is clearly seen from Fig.\,\ref{fig_EsymFSUre} that the applicable region of the conventional expansion (smaller than about $2\rho_0$) is much smaller than
that of the expansion (\ref{tt1}) (up to about $5\rho_0$), at both sub- and supra-saturation densities.
These result show that the reconstruction (\ref{tt1}) with a reasonable $\Ts$ to order $\nu_{\rm{sym}}^3$
can well reconstruct the symmetry energy predicted by the RMF/FSUGold model. Moreover, using the algebraic model for $\nu_{\rm{sym}}$ we can obtain a very similar reconstruction by adjusting the parameter $\Ts$.

\section{Predicting Nuclear Symmetry Energy at $\rho_0\lesssim\rho\lesssim3\rho_0$ using its characteristics at $\rho_0$}\label{S3}
In this section, using the auxiliary-function-based reconstruction with the condition (\ref{fit-sk}) for determining the $\Ts$ we investigate the symmetry energy at supra-saturation densities in the range of $\rho_0\lesssim\rho\lesssim3\rho_0$ based on its characteristic parameters at $\rho_0$.

\subsection{Results of Monte Carlo Simulations}
For our Monte Carlo simulations, the following three test sets with the specified characteristic parameters of $E_{\rm{sym}}(\rho)$ at $\rho_0$ are considered:
\begin{enumerate}[leftmargin=*,label=\Roman*.]
\item The $E_{\rm{sym}}(\rho)$ is expanded to order $\nu_{\rm{sym}}^2$ or order $\chi^2$ with $-300\,\rm{MeV}\leq K_{\rm{sym}}\leq 0\,\rm{MeV}$\,\cite{Xie20}.
\item The $E_{\rm{sym}}(\rho)$ is expanded to order $\nu_{\rm{sym}}^3$ or order $\chi^3$ with $-300\,\rm{MeV}\leq K_{\rm{sym}}\leq 0\,\rm{MeV}, 0\,\rm{MeV}\leq J_{\rm{sym}}\leq2000\,\rm{MeV}$.
\item The $E_{\rm{sym}}(\rho)$ is expanded to order $\nu_{\rm{sym}}^3$ or order $\chi^3$ with $K_{\rm{sym}}$ and $J_{\rm{sym}}$ given by the following intrinsic relations imposed by the unbound nature of pure neutron matter (PNM)\,\cite{Cai20},
\begin{align}
K_{\rm{sym}}\approx&K_0\left(1-\frac{1}{3}\frac{K_0}{L}+\frac{1}{2}\frac{J_0}{K_0}\frac{L}{K_0}\right),\label{IC1}\\
J_{\rm{sym}}\approx&\frac{2K_0^3}{3L^2}\left(1-\frac{3L}{K_0}\right)+\frac{I_0L}{3K_0}\notag\\
&+\left(\frac{2K_0K_{\rm{sym}}}{L}-J_0\right)
\left(1+\frac{J_0L}{K_0^2}-\frac{K_{\rm{sym}}}{K_0}\right).\label{IC2}
\end{align}
\end{enumerate}
Here $K_0,J_0,I_0$ are the incompressibility, skewness and kurtosis of the EOS of SNM, appeared in the expansion $E_0(\rho)\approx E_0(\rho_0)+2^{-1}K_0\chi^2+6^{-1}J_0\chi^3+24^{-1}I_0\chi^4+\cdots$.
These sets are denoted as ``set I'', ``set II'', and ``set III'', respectively.
For the main physical demonstrations, we adopt in the Monte Carlo simulations $S\approx32\pm4\,\rm{MeV}, L\approx60\pm30\,\rm{MeV}$\,\cite{LiBA13}, $K_0\approx240\pm40\,\rm{MeV}$\,\cite{Garg18,You99,Shl06,Che12,Col14}, $J_0\approx-300\pm200\,\rm{MeV}$\,\cite{Cai17x}, and $I_0\approx0\pm2000\,\rm{MeV}$, respectively.

In Fig.\,\ref{fig_Esymu-set-ABC}, the symmetry energy $E_{\rm{sym}}(\rho)$ from simulations adopting the exponential model (abbreviated as ``exp'' in the figure) is shown for the test set I (panel (a)), II (panel (b)), and III (panel (c)), respectively.
The statistical error shown in the figure, i.e., $E_{\rm{sym}}(\rho_{\rm{f}})\pm\sigma[{E_{\rm{sym}}(\rho_{\rm{f}})}]$ where $\rho_{\rm{f}}$ is a reference density, is calculated via
\begin{equation}
\sigma[f]=\sqrt{\langle f^2\rangle-\langle f\rangle^2}=
\sqrt{
\frac{1}{n}\sum_{i=1}^nf^{(i),2}-\left(\frac{1}{n}\sum_{i=1}^nf^{(i)}\right)^2
},
\end{equation}
here $f^{(i)}$ is the reconstructed symmetry energy from the $i$th independent run of the Monte Carlo samplings, and in our simulations $n=10^4$ is used. 
In fact, the uncertainty of $E_{\rm{sym}}(\rho_{\rm{f}})$ is determined mainly by the uncertainties of the characteristics $L, K_0$, etc., and has tiny dependence on $n$.
Specifically, the $\Ts$ parameter in set I is found to be about $\Ts\approx 1.67\pm0.56$, while that in set II (set III) is found to be about $\Ts\approx 1.41\pm0.88$ ($1.74\pm0.81$).
The $\nu_{\rm{sym}}^3$ order contribution introduces about $-11$\% (4\%) effect on $\Ts$ in set II (set III) compared with that in set I, with the latter truncated at order $\nu_{\rm{sym}}^2$.

\renewcommand*\tablename{\small Tab.}
\begin{table}[h!]
\centering
\begin{tabular}{c||c|c||c|c}
\hline $E_{\rm{sym}}(\rho)$
& $2\rho_0$ [exp] & $3\rho_0$ [exp]&$2\rho_0$ [alge]& $3\rho_0$ [alge]
\\
\hline\hline
AUX [I] & $44.8\pm5.9$ &$48.6\pm8.9$&$47.3\pm6.5$&$53.9\pm13.4$\\ \hline
CON[I] &$43.7\pm8.4$ &$36.0\pm23.7$ &$45.9\pm7.9$&$42.9\pm21.8$ \\ \hline
\hline 
AUX [II]&$43.3\pm6.4$&$51.8\pm18.0$&$43.5\pm6.5$&$52.1\pm18.7$\\ \hline
CON  [II]&$44.9\pm8.1$&$72.0\pm36.3$&$44.8\pm7.5$&$70.4\pm33.7$ \\ \hline
\hline
AUX [III]&$43.7\pm7.6$&$44.7\pm22.3$&$44.4\pm9.2$&$43.2\pm31.2$\\ \hline
CON  [III]&$40.3\pm10.5$&$13.4\pm46.0$&$42.8\pm10.3$&$26.7\pm42.7$ \\ \hline
\end{tabular}%
\caption{Nuclear symmetry energies at $2\rho_0$ and $3\rho_0$ from the auxiliary-function-based (abbreviated as ``AUX'') and the conventional (abbreviated as ``CON'') expansions in the two models, unit:\,MeV, ``exp'' and ``alge'' abbreviate for the exponential and the algebraic models.}  
\label{tab_Esym2u3u}
\end{table}

Several interesting features are demonstrated in Fig.\,\ref{fig_Esymu-set-ABC}. Firstly, below about $1.5\rho_0$ the auxiliary-function-based and the conventional expansions give almost identical results. At higher densities, however, 
changing from the test set I to set III, the result from the auxiliary-function-based approach is stable and always has smaller error bars compared to that from the conventional expansion. In another words, 
as the truncation orders and/or the empirical values for $K_{\rm{sym}}$ and $J_{\rm{sym}}$ change (panels (b) and (c)), the prediction on the symmetry energy at supra-saturation densities from the auxiliary-function-based expansion is more stable than that from the conventional expansion approach. These features are more quantitatively demonstrated in the 2nd and 3rd columns of Tab.\,\ref{tab_Esym2u3u}. The 4th and 5th columns are results using the algebraic model  to be discussed in Section \ref{Ind}. In particular, the $E_{\rm{sym}}(2\rho_0)$ in the auxiliary-function-based expansion changes from 44.8\,MeV in set I to 43.3\,MeV (43.7\,MeV) in set II (set III), generating a difference of about $-3.3$\% ($-2.5$\%). Very similarly, the $E_{\rm{sym}}(3\rho_0)$ changes from 48.6\,MeV in set I to 51.8\,MeV (44.7\,MeV) in set II (set III),
and the relative change is found to be about 6.6\% ($-8.0$\%). On the other hand, the $E_{\rm{sym}}(3\rho_0)$ changes from 36.0\,MeV in the conventional expansion in set I to 72.0\,MeV (13.4\,MeV) in set II (set III), with the latter inducing a difference of about 100\% ($-62.7$\%).

Secondly, noticing that the test set I is at $\nu_{\rm{sym}}^2$ or $\chi^2$ orders while the test set II and set III are both at  $\nu_{\rm{sym}}^3$ or $\chi^3$ orders, the results shown in Fig.\,\ref{fig_Esymu-set-ABC} and Tab.\,\ref{tab_Esym2u3u} indicate that the higher order contributions from $\nu_{\rm{sym}}^3$ are relatively small in the auxiliary-function-based reconstruction. In fact, as already mentioned earlier, the function $\nu_{\rm{sym}}$ itself can generate higher order terms in $\chi$ like $\chi^3$ and $\chi^4$, etc., even when the symmetry energy is truncated apparently at order $\nu_{\rm{sym}}^2$. 
It is thus not surprising that the predicted symmetry energy at supra-saturation densities from the auxiliary-function-based reconstruction either to order $\nu_{\rm{sym}}^2$ or to order $\nu_{\rm{sym}}^3$ looks very similar, since the effects from the characteristic parameter $J_{\rm{sym}}$ and even higher order contributions are modeled effectively with the help of $\nu_{\rm{sym}}$ in the set I by adaptively adjusting the $\Ts$ parameter, and these terms are included directly in the set II and set III simulations.
On the other hand, compared to set I, the $E_{\rm{sym}}(\rho)$ from the set II from the conventional expansion quickly becomes stiffer at supra-saturation densities since a positive $J_{\rm{sym}}$ is used (which is absent in the set I). Similar phenomenon occurs when the intrinsic relations for $J_{\rm{sym}}$ and $K_{\rm{sym}}$ are used in set III, i.e., the $E_{\rm{sym}}(\rho)$ quickly becomes softer at densities $\rho\gtrsim2.5\rho_0$.

Finally, the large uncertainty band $\sigma[E_{\rm{sym}}(\rho_{\rm{f}})]$ needs some more explanations and the relevant context. Firstly, the uncertainty is mostly coming from the uncertainties of $S, L$, $K_{\rm{sym}}$ and  $J_{\rm{sym}}$. Of course, for each set of fixed values of these parameters, the $E_{\rm{sym}}(\rho_{\rm{f}})$ is fixed at a unique value. While in our Monte Carlo simulations that generate randomly the initial values of these parameters within their $1\sigma$ uncertainty range subject to the condition of Eq. (\ref{fit-sk}) for fixing the $\Ts$, the resulting $E_{\rm{sym}}(\rho_{\rm{f}})$ at any given density $\rho_{\rm{f}}$ is approximately a Gaussian distribution as one expects statistically. Practically, in fact, in the nowadays widely used Bayesian statistical inference of model parameters directly from data of the observations, the posterior probability distribution functions (PDFs) of the model parameters are typically characterized by their most probable (the peak of the PDF which is of course different from the mean when the PDF is asymmetric) value and a variable confidence boundary, e.g., 68\% ($1\sigma$) or 95\% ($2\sigma$). Correspondingly, functions reconstructed using model parameters extracted in such approach are generally described with the most probable (or mean) value together with selected confidence boundaries, see, e.g., examples given in Refs. \cite{Xie19,Xie20}. Our presentations should be understood in this context. 
Secondly, although the uncertainties of $E_{\rm{sym}}(\rho_{\rm{f}})$ reconstructed using the auxiliary functions are significantly reduced compared to those in the traditional approach using the Taylor expansion, 
as indicated clearly in Fig.\,\ref{fig_Esymu-set-ABC}, Tab.\,\ref{tab_Esym2u3u} and also Fig.\,\ref{fig_Esym_exp_alge}, the uncertainties of $E_{\rm{sym}}(\rho_{\rm{f}})$ at $(2-3)\rho_0$ are still very large, motivating/requiring many ongoing/future works in this field.

\subsection{Understanding the Dependence of High-Density Symmetry Energy $E_{\textmd{sym}}(\rho)$ on its  Characteristics at $\rho_0$}
While in the conventional expansion of Eq.\,(\ref{EXPEsym}) the dependences of $E_{\rm{sym}}(\rho)$ on its characteristic parameters at $\rho_0$ are obvious by definition, these dependences are no longer obvious in the auxiliary-function-based expansions because of the convolutions. Nevertheless, they can be mathematically analyzed rigorously by calculating the relevant derivatives. 
For example, the dependences of $E_{\rm{sym}}(\rho_{\rm{f}})$ at a reference density $\rho_{\rm{f}}$ on $K_{\rm{sym}}$ and $J_{\rm{sym}}$ can be analyzed within the auxiliary-function-based expansion with the exponential model to order $\nu_{\rm{sym}}^3$ using the following derivatives,
\begin{align}
\frac{\partial E_{\rm{sym}}(\rho_{\rm{f}})}{\partial K_{\rm{sym}}}=&
\frac{1}{2}\nu_{\rm{sym}}^{\rm{f},2}\left(1+3\Ts\nu_{\rm{sym}}^{\rm{f}}\right)\notag\\
&\times\left[1-\left(\frac{\nu_{\rm{sym}}^{\rm{low}}}{\nu_{\rm{sym}}^{\rm{f}}}\right)^2\left(\frac{1+3\Ts\nu_{\rm{sym}}^{\rm{low}}}{1+3\Ts\nu_{\rm{sym}}^{\rm{f}}}\right)\left(\frac{\Upsilon_{\rm{f}}}{\Upsilon_{\rm{low}}}\right)\right]
,\label{Ek}\\
\frac{\partial E_{\rm{sym}}(\rho_{\rm{f}})}{\partial J_{\rm{sym}}}=&\frac{1}{6}\nu_{\rm{sym}}^{\rm{f},3}\times
\left[1-\left(\frac{\nu_{\rm{sym}}^{\rm{low}}}{\nu_{\rm{sym}}^{\rm{f}}}\right)^3\left(\frac{\Upsilon_{\rm{f}}}{\Upsilon_{\rm{low}}}\right)\right],\label{Ej}
\end{align}
where the superscripts/subscripts ``f'' and ``low'' are for $\rho_{\rm{f}}$ and $\rho_{\rm{low}}\approx0.05\,\rm{fm}^{-3}$ (see the fitting scheme (\ref{fit-sk})), respectively.
The function $\Upsilon(\rho)$ in the above two equations is given by
\begin{align}
\Upsilon(\rho)
=&\frac{3}{2}L\nu_{\rm{sym}}^2+\frac{3}{2}\left(K_{\rm{sym}}+4L\Ts\right)\nu_{\rm{sym}}^3\notag\\
&+\frac{\partial\nu_{\rm{sym}}}{\partial\Ts}\Bigg[L+\left(K_{\rm{sym}}+3L\Ts\right)\nu_{\rm{sym}}\notag\\
&+\frac{1}{2}\left(J_{\rm{sym}}+9K_{\rm{sym}}\Ts+18L\Ts^2\right)\nu_{\rm{sym}}^2
\Bigg].\end{align}
The corrections in the square brackets in (\ref{Ek}) and (\ref{Ej}) come from the dependence of the $\Ts$ parameter on the curvature $K_{\rm{sym}}$ and the skewness $J_{\rm{sym}}$ of the symmetry energy, i.e., $\partial\Ts/\partial K_{\rm{sym}}$ and $\partial\Ts/\partial J_{\rm{sym}}$.
By adopting the empirical values of $J_{\rm{sym}}\approx300\,\rm{MeV},K_{\rm{sym}}\approx-80\,\rm{MeV},L\approx60\,\rm{MeV}$ and the $\Ts\approx1.74$ (set III), respectively, we immediately find $\partial E_{\rm{sym}}(3\rho_0)/\partial K_{\rm{sym}}\approx0.0495$ and $\partial E_{\rm{sym}}(3\rho_0)/\partial J_{\rm{sym}}\approx0.0026$.
On the other hand, we have in the conventional reconstruction that ${\partial E_{\rm{sym}}(\rho_{\rm{f}})}/{\partial K_{\rm{sym}}}=\chi_{\rm{f}}^2/2$ and ${\partial E_{\rm{sym}}(\rho_{\rm{f}})}/{\partial J_{\rm{sym}}}=\chi_{\rm{f}}^3/6$, and thus $\partial E_{\rm{sym}}(3\rho_0)/\partial K_{\rm{sym}}\approx0.2223$ and $\partial E_{\rm{sym}}(3\rho_0)/\partial J_{\rm{sym}}\approx0.0494$, respectively.
The small value of $\partial E_{\rm{sym}}(3\rho_0)/\partial J_{\rm{sym}}\approx0.26$\% demonstrates again that the dependence of $E_{\rm{sym}}(\rho_{\rm{f}})$ with $\rho_{\rm{f}}\lesssim3\rho_0$ on the skewness $J_{\rm{sym}}$ is weak in the auxiliary-function-based reconstruction approach, e.g., $\delta J_{\rm{sym}}\partial E_{\rm{sym}}(3\rho_0)/\partial J_{\rm{sym}}\approx1.3\,\rm{MeV}$ if $\delta J_{\rm{sym}}\approx 500\,\rm{MeV}$. The same uncertainty of $J_{\rm{sym}}$ leads however to an uncertainty of about 24.7\,MeV for $E_{\rm{sym}}(3\rho_0)$ in the conventional reconstruction, which is about 19 times larger than the value in the auxiliary-function-based approach. We found similar conclusions with the algebraic model.

\renewcommand*\figurename{\small Fig.}
\begin{figure}[h!]
\centering
\includegraphics[width=6.8cm]{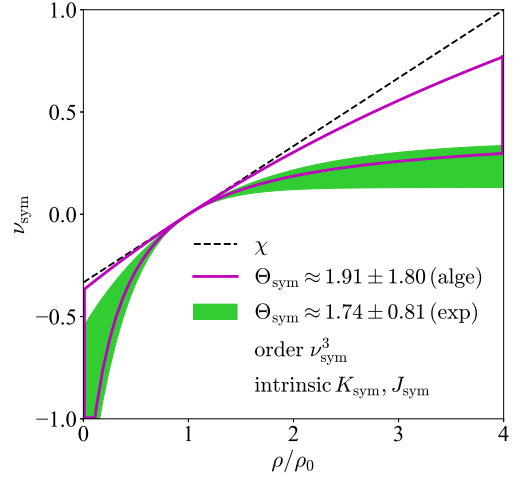}
 \caption{(Color Online). The expansion variable $\nu_{\rm{sym}}$ as a function of density within the two auxiliary models.}\label{fig_nusym_exp_alge}
\end{figure}

The main reason for the weak dependence of high-density $E_{\rm{sym}}(\rho)$ on $J_{\rm{sym}}$ in the auxiliary-function-based reconstruction is that the expansion variable $\nu_{\rm{sym}}$ converges much faster than $\chi$. In Fig.\,\ref{fig_nusym_exp_alge} we show the density dependence of $\nu_{\rm{sym}}$. It is clearly seen that the $\nu_{\rm{sym}}$ quickly approaches a small value significantly below $\chi$ at supra-saturation densities, guaranteeing the faster convergence of the auxiliary function approach. More quantitatively, we find that at $\rho=3\rho_0$ (with $\chi=2/3,\chi^2=4/9$ and $\chi^3=8/27$) that $\nu_{\rm{sym}}\approx0.186,\nu_{\rm{sym}}^2\approx0.034$ and $\nu_{\rm{sym}}^3\approx0.006$ in the exponential model, while $\nu_{\rm{sym}}\approx0.288,\nu_{\rm{sym}}^2\approx0.083$ and $\nu_{\rm{sym}}^3\approx0.024$ in the algebraic model.
These numbers show again that the contribution from order $\nu_{\rm{sym}}^3$ is relatively small in the auxiliary-function-based approach, and thus has little impact.

It is necessary to point out that although the $\nu_{\rm{sym}}$ factor becomes small even approaching zero at supra-saturation densities, its value deviates from zero faster than $\chi$ at sub-saturation densities.
This means that the converges of the reconstruction of the symmetry energy using the auxiliary function approach becomes worse at sub-saturation densities.
Thus, it may be practical and important to consider using one auxiliary function at supra-saturation and another one at sub-saturation densities, or simply use the $\Ts$ as a free parameter such that the expansion factor $\nu_{\rm{sym}}$ converges at both small and large densities. These issues are among the topics of our future studies.

\subsection{Independence of The High-density $E_{\textmd{sym}}(\rho)$ \\ on The Auxiliary Function Selected}\label{Ind}
In order to investigate possible model-dependences of the reconstructed high-density symmetry energy due to the function $\nu_{\rm{sym}}(\chi,\Ts)$ or $\Ps$ selected, we have plotted curves similar to those shown in Fig.\,\ref{fig_Esymu-set-ABC} by adopting the algebraic model (\ref{nnn-1}) instead. The resulting symmetry energies at $2\rho_0$ and $3\rho_0$ are listed in the 4th and 5th columns of Tab.\,\ref{tab_Esym2u3u}. As the test set I, set II and set III themselves are very similar either in the exponential model or in the algebraic model, in the following we focus on the results obtained in the test set III, where the intrinsic correlations among the characteristics of $E_{\rm{sym}}(\rho)$ at $\rho_0$ are adopted for the simulations.

\begin{figure}[h!]
\centering
\includegraphics[width=6.5cm]{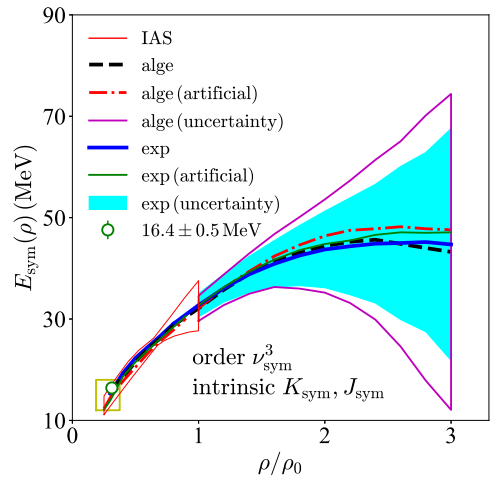}
 \caption{(Color Online). Symmetry energy reconstructed via the auxiliary functions in the exponential and algebraic models with the test set III (adopting the intrinsic correlation between $K_{\rm{sym}}$ and $J_{\rm{sym}}$ imposed by the unbound nature of PNM).}\label{fig_Esym_exp_alge}
\end{figure}

In Fig.\,\ref{fig_Esym_exp_alge}, the density dependence of nuclear symmetry energy from $0.3\rho_0$ to $3\rho_0$ with its $1\sigma$ uncertainty band is shown with the exponential and algebraic auxiliary function, respectively. The $\Ts$ parameter for the algebraic model in the set III is found to be about $\Ts\approx1.91\pm1.80$ while that in the exponential model is $\Ts\approx1.74\pm 0.81$.
It is seen clearly that the $E_{\rm{sym}}(\rho)$ obtained from the two models (blue-solid and black-dashed) behave very similarly albeit with slightly different error bands (cyan and magenta), indicating the reconstruction is effective and largely independent of the auxiliary function used. 

It is interesting to note that the $E_{\rm{sym}}(\rho)$ at sub-saturation densities is found consistent with the result from analyzing the isobaric-analog-state (IAS) data (indicated by the red-solid curve)\,\cite{Dan14}. Moreover, the reference symmetry energy $E_{\rm{sym}}(0.05\,\rm{fm}^{-3})=16.4\pm 0.5$\,MeV (green circle) we used in fixing the $\Ts$ parameter is also consistent with the IAS result. 

In order to investigate how the predicted high-density $E_{\rm{sym}}(\rho)$ may depend on the scheme of Eq.\ (\ref{fit-sk}) for fixing the $\Ts$ parameter, we have done a test by artificially extending $\rho_{\rm{low}}$ to $0.03$-$0.06\,\rm{fm}^{-3}$ and taking correspondingly $E_{\rm{sym}}(\rho_{\rm{low}})=12$-$18\,\rm{MeV}$, as shown in Fig.\,\ref{fig_Esym_exp_alge} by the yellow box. 
The symmetry energies thus obtained from the two models still show very similar behavior in the density region of $0.5\rho_0$-$3\rho_0$, as shown by the green-solid and red-dash-dotted lines (indicated by ``artificial'' in the parentheses).  More quantitatively, we have now $E_{\rm{sym}}^{\rm{exp}}(2\rho_0)\approx44.8\pm8.1\,\rm{MeV}$ and $E_{\rm{sym}}^{\rm{alge}}(2\rho_0)\approx 46.4\pm9.1\,\rm{MeV}$, as well as $E_{\rm{sym}}^{\rm{exp}}(3\rho_0)\approx 47.1\pm22.7\,\rm{MeV}$ and $E_{\rm{sym}}^{\rm{alge}}(3\rho_0)\approx47.6\pm31.7\,\rm{MeV}$, respectively. Therefore, although the determination scheme for $\Ts$ has been changed, its effect on the predicted high-density $E_{\rm{sym}}(\rho)$ is minor, e.g., $E_{\rm{sym}}^{\rm{exp}}(2\rho_0)$ changes from 43.7\,MeV to 44.8\,MeV, and $E_{\rm{sym}}^{\rm{alge}}(2\rho_0)$ changes from 44.4\,MeV to 46.4\,MeV, etc, showing quantitatively the stability of the prediction once again.

 \renewcommand*\figurename{\small Fig.}

Despite of the fact that very similar $E_{\rm{sym}}(\rho_{\rm{f}})$ functions with $\rho_{\rm{f}}\lesssim3\rho_0$ are obtained by using either the exponential or the algebraic model, 
they are not enough to prove model independence of our results. Nevertheless, one can argue that all the models $\nu_{\rm{sym}}(\chi,\Ts)$ at low-density scales of $\chi$ are effectively equivalent, i.e., $\nu_{\rm{sym}}(\chi,\Ts)\approx\chi+\mbox{``}\rm{corrections}\mbox{''}$, and the higher order contributions (characterized by characteristics like $J_{\rm{sym}}$) could be effectively absorbed into the lower-order coefficients via the parameter $\Ts$.
These higher-order terms are expected to have little impact on the low-density scales of $\chi$, which is the basic consideration of the auxiliary-function-based reconstruction. However, to verify these expectations quantitatively, 
further investigations using  more different forms of $\Ps$ will be extremely useful.

\subsection{High-density Trend and \\Limit of Nuclear Symmetry Energy}
It is also interesting to note from Fig.\,\ref{fig_Esym_exp_alge} that the mean value of $E_{\rm{sym}}(\rho)$ from adopting the intrinsic relations (\ref{IC1}) and (\ref{IC2}) either in the exponential model or in the algebraic model shows a decreasing or flattening trend starting around $2\rho_0$-$3\rho_0$. 
Since the intrinsic relations used in the test set III are obtained from a model-independent manner by considering only the unbound nature of PNM\,\cite{Cai20}, the $E_{\rm{sym}}(\rho)$ with $\rho\lesssim3\rho_0$ from the auxiliary-function-based reconstruction is also expected to be rather general.

Since the high-density behavior of $E_{\rm{sym}}(\rho)$ has long been debated in nuclear physics, see, e.g., Refs. \cite{Xiao,Feng12,Xie14,MSU,Liu} and it has significant implications for neutron stars \cite{Kut93,Kub99,Kut00,wen,NPnews,YZhou19-a}, a few more comments are necessary here. As noticed earlier, the curvature $K_{\rm{sym}}\approx-118\,\rm{MeV}$ is found from the relation (\ref{IC1}) by adopting the empirical values of $K_0\approx240\,\rm{MeV}, J_0\approx-300\,\rm{MeV}$ and $L\approx60\,\rm{MeV}$. It is very close to the $K_{\rm{sym}}\approx-120^{+80}_{-100}\,\rm{MeV}$ obtained from a recent Bayesian analysis of several observables of neutron stars\,\cite{Xie20}. 
The negative value of $K_{\rm{sym}}$ naturally leads to the bending down or flattening of the symmetry energy at some supra-saturation density unless the skewness $J_{\rm{sym}}$ is extremely large and positive. But as we have discussed earlier, the latter plays little role in determining the $E_{\rm{sym}}(\rho)$ around $2\rho_0$-$3\rho_0$ in the auxiliary function approach. Thus, it is not surprising to see the flattening or decreasing trend of $E_{\rm{sym}}(\rho)$ starting around $2\rho_0$-$3\rho_0$.  However, it is too early to conclude that the symmetry energy definitely decreases beyond this density region as the uncertainty is still very large at $\rho\approx3\rho_0$.

Moreover, some quantitative comparisons of the $E_{\rm{sym}}(\rho)$ values around $2\rho_0$-$3\rho_0$ reconstructed here with those from other approaches are useful. 
Combining the $E_{\rm{sym}}(\rho)$ reconstructed from the two models for $\Ps$ with the test set III, we find the effective central value of $43\,\rm{MeV}\lesssim E_{\rm{sym}}(3\rho_0)\lesssim54\,\rm{MeV}$ and $43\,\rm{MeV}\lesssim E_{\rm{sym}}(2\rho_0)\lesssim48\,\rm{MeV}$, respectively. As shown in Table \ref{tab_Esym2u3u}, similar values are obtained with the other two test sets. Interestingly, the reconstructed $E_{\rm{sym}}(2\rho_0)$ value is consistent with its current fiducial value and predictions of the state-of-the-art nuclear many-body theories. In particular, eight earlier independent analyses of some heavy-ion reaction data and neutron star properties\,\cite{Xie20} gave a fiducial value of about 47\,MeV for $E_{\rm{sym}}(2\rho_0)$. It is consistent with the very recent prediction of $E_{\rm{sym}}(2\rho_0) \leq 53.2$ MeV based on a nuclear energy density functional theory \cite{PKU-Meng}, $E_{\rm{sym}}(2\rho_0) \approx 46 \pm 4$ MeV \cite{Diego} based on the Quantum Monte Carlo calculations, and $E_{\rm{sym}}(2\rho_0) \approx 45 \pm 3$ MeV \cite{Ohio20} from the latest many-body perturbation theory calculation using consistent nucleon-nucleon and three-nucleon interactions up to fourth order in the chiral effective field expansion.

Finally, we have not attempted to explore the symmetry energy at densities above $3\rho_0$ in the auxiliary-function-based approach. The reason is twofold. Firstly, as the density increases beyond $3\rho_0$, the $E_{\rm{sym}}(\rho)$ from the exponential and algebraic models shows systematic differences (either in set I, set II or set III), indicating a serious dependence on the auxiliary function $\Ps$. Secondly, a hadron-quark phase transition and non-nucleonic degrees of freedom are likely to appear above $3\rho_0$ and hence there is no basic need there to define a nucleonic symmetry energy, see mode detailed discussions in, e.g., Ref.\,\cite{XieLi-PRC21}. 

\section{Summary and outlook}\label{S4}
In summary, by adopting the auxiliary-function-based expansion for nuclear symmetry energy at suprasaturation densities, one can effectively incorporate contributions from its higher order characteristics at saturation density.
The symmetry energy in the density region of $\rho_0\lesssim\rho\lesssim3\rho_0$ is found to be stable irrespective of the truncation order in the approach and/or the empirical values for the higher order characteristics like $K_{\rm{sym}}$ and $J_{\rm{sym}}$ adopted. The reconstructed symmetry energy at suprasaturation densities from the new approach has smaller error bars compared to that from using the conventional $\chi$-expansion. 
Moreover, the symmetry energy $E_{\rm{sym}}(\rho)$ at densities $2\rho_0\lesssim\rho\lesssim3\rho_0$ is found to be flat and trending down as the density increases. 

The auxiliary function approach is found to converge much faster than the conventional $\chi$-expansion in the density region of $\rho_0\lesssim\rho\lesssim3\rho_0$.
In principle, it can be applied not only to expanding the symmetry energy but also the EOS of SNM. The conventional expansion of the latter based on $\chi$ suffers from similar shortcomings as in expanding the symmetry energy. Moreover, the auxiliary function approach may also be used to study simultaneously the isospin and density dependences of superdense neutron-rich matter by reforming the expansion of its EOS in terms of the isospin asymmetry $\delta^2$ via a similar transform $\Omega(\delta,\Delta)$ similar to the $\Ps$, where $\Delta$ is a parameter of $\Omega$ similar to the $\Ts$ used in the $\Ps$ function. Applications of such approach may be useful for extracting more accurately the EOS of superdense neutron-rich matter from structures and collision prooducts of both heavy nuclei in terrestrial laboratories and neutron stars in heaven. 

\section*{Acknowledgement}
This work is supported in part by the U.S. Department of Energy, Office of Science, under Award Number DE-SC0013702, the CUSTIPEN (China-U.S. Theory Institute for Physics with Exotic Nuclei) under the US Department of Energy Grant No. DE-SC0009971.

\appendix

\renewcommand\theequation{a\arabic{equation}}
\section{Proof on Equivalence between $\Phi=0$ of Eq.\,(\ref{def_PHI}) and Expansion of $E_{\mathrm{sym}}(\rho)\approx S+L\nu_{\mathrm{sym}}\approx S+L\chi+2^{-1}K_{\mathrm{sym}}\chi^2+\mathcal{O}(\chi^3)$}

In this appendix, we prove that the expansion of the symmetry energy in the auxiliary-function-based reconstruction at the truncation order of $\nu_{\rm{sym}}$, i.e., $E_{\mathrm{sym}}(\rho)\approx S+L\nu_{\mathrm{sym}}$, to the conventional order of $\chi^2$,  namely $E_{\rm{sym}}(\rho)\approx S+L\chi+K_{\rm{sym}}\chi^2/2+\mathcal{O}(\chi^3)$ is equivalent to condition $\Phi=0$ with $\Phi$ defined in (\ref{def_PHI}).

Expanding $\Ps$ around $\chi\approx0$ gives
\begin{align}
\Ps\approx &\Pso+\left.\frac{\partial \Pi_{\rm{sym}}}{\partial\chi}\right|_{\chi=0}\cdot\chi\notag\\
&+\left.\frac{1}{2}\frac{\partial^2\Pi_{\rm{sym}}}{\partial\chi^2}\right|_{\chi=0}\cdot\chi^2+\mathcal{O}(\chi^3).
\end{align}
The expansion of the symmetry energy $E_{\mathrm{sym}}(\rho)\approx S+L\nu_{\mathrm{sym}}$ then becomes (where  $\widetilde{\nu}_{\rm{sym}}=\Ps-\Pso$),
\begin{align}\label{app}
E_{\rm{sym}}(\rho)\approx&S+\left.\frac{L}{3\rho_0}\frac{\partial\rho}{\partial\Pi_{\rm{sym}}}\right|_{\chi=0}\cdot\widetilde{\nu}_{\rm{sym}}\notag\\
=&S+L\chi+\left.\frac{3\rho_0L}{2}\frac{\partial^2\Pi_{\rm{sym}}/\partial\rho^2}{\partial\Pi_{\rm{sym}}/\partial\rho}\right|_{\chi=0}\cdot\chi^2.
\end{align}

By using the basic relations between derivatives, i.e.,
\begin{align}
\frac{\partial\Pi_{\rm{sym}}}{\partial\rho}=&\left(\frac{\partial\rho}{\partial\Pi_{\rm{sym}}}\right)^{-1},~~\\
\frac{\partial^2\Pi_{\rm{sym}}}{\partial\rho^2}=&-\left(\frac{\partial\Pi_{\rm{sym}}}{\partial\rho}\right)^3\cdot\frac{\partial^2\rho}{\partial\Pi_{\rm{sym}}^2},
\end{align}
we can rewrite the last term in (\ref{app}) as
\begin{equation}
-\frac{3\rho_0L}{2}\left[\left(\frac{\partial\Pi_{\rm{sym}}}{\partial\rho}\right)^2\cdot\frac{\partial^2\rho}{\partial\Pi_{\rm{sym}}^2}\right]_{\chi=0}\cdot\chi^2,
\end{equation}
and then make it to be equal to $K_{\rm{sym}}\chi^2/2$, leading to
\begin{equation}
K_{\rm{sym}}+3\rho_0L\left[\left(\frac{\partial\Pi_{\rm{sym}}}{\partial\rho}\right)^2\cdot\frac{\partial^2\rho}{\partial\Pi_{\rm{sym}}^2}\right]_{\chi=0}=0,
\end{equation}
which is just the condition $\Phi=0$.


\begin{references}

\bibitem{Lat00} J.M. Lattimer, M. Prakash, Phys. Rep. \textbf{333}, 121 (2000).

\bibitem{ditoro} V. Baran, M. Colonna, V. Greco, and M.Di Toro, Phys. Rep. \textbf{410}, 335 (2005).

\bibitem{Steiner05} A.W. Steiner, M. Prakash, J.M. Lattimer, and P.J. Ellis, Phys. Rep., 410, 325 (2005).

\bibitem{LCK08} B.A. Li, L.W. Chen, and C.M. Ko, Phys. Rep. \textbf{464}, 113 (2008).

\bibitem{Ditoro10}M. Di Toro, V. Baran, M. Colonna, V. Greco, J. Phys. G: Nucl. Part. Phys. {\bf 37}, 083101 (2010).

\bibitem{Chen11} L.W. Chen, Sci. China Phys. Mech. Astron. \textbf{54}, suppl.1, s124 (2011).

\bibitem{Trau12} W. Trautmann, H.H. Wolter, Int. J. Mod. Phys. E \textbf{21}, 1230003 (2012).

\bibitem{Tsang12} M.B. Tsang \textit{et al.}, Phys. Rev. C \textbf{86}, 015803 (2012).

\bibitem{Tesym} B.A. Li, \`{A}. Ramos, G. Verde, I. Vida\~{n}a (Eds.), \textit{Topical issue on nuclear symmetry energy}, Euro. Phys. J. A \textbf{50}, No.2 (2014).

\bibitem{Bal16} M. Baldo and G.F. Burgio, Prog. Part. Nucl. Phys. \textbf{91}, 203 (2016).

\bibitem{Oer17} M. Oertel, M. Hempel, T. Klahn, and S. Typel, Rev. Mod. Phys.  \textbf{89}, 015007 (2017).

\bibitem{BALI19}B.A. Li, P.G. Krastev, D.H. Wen, and N.B. Zhang, Euro. Phys. J. A, {\bf 55}, 117 (2019).

\bibitem{Steiner} A.W. Steiner, J.M. Lattimer, and E.F. Brown, APJ, {\bf 722}, 33 (2010).

\bibitem{Zhang18} N.B. Zhang, B. A. Li, and J. Xu, APJ, {\bf 859}, 90 (2018).

\bibitem{MM1} J. Margueron, R. Casali, F. Gulminelli, Phys. Rev. C \textbf{96}, 065805 (2018);
\textit{ibid}, \textbf{96}, 065806 (2018).


\bibitem{Xie19}W.J. Xie and B. A. Li, APJ, {\bf 883}, 174 (2019).

\bibitem{France1} N. Baillot d'Etivaux, S. Guillot, J. Margueron, N.A. Webb, M. Catelan, and A. Reisenegger, Astrophys. J., {\bf 887}, 48 (2019).

\bibitem{Sof}Sofija Anti\'c, Debarati Chatterjee, Thomas Carreau, Francesca Gulminelli, J. Phys. G: Nucl. Part. Phys. {\bf 46}, 065109 (2019).

\bibitem{Xie20}W.J. Xie and B.A. Li, ApJ. \textbf{899}, 4 (2020).

\bibitem{Con2} Márcio Ferreira, Renan Câmara Pereira, Constança Providência, Phys. Rev. D {\bf 101}, 123030 (2020).

\bibitem{burg} G. F. Burgio and I. Vida\~{n}a, Universe {\bf 6(8)}, 119 (2020).

\bibitem{Tsang} C.Y. Tsang, M.B. Tsang, P. Danielewicz, W.G. Lynch, F.J. Fattoyev, Phys. Rev. C {\bf 102}, 045808 (2020).

\bibitem{Bis21}Bhaskar Biswas, Prasanta Char, Rana Nandi, Sukanta Bose,  arXiv:2008.01582 (2021).

\bibitem{LiBA13} B.A. Li and X. Han, Phys. Lett. \textbf{B727}, 276 (2013).

\bibitem{YZhou19}Y. Zhou, L.W. Chen, and Z. Zhang, Phys. Rev. D \textbf{99}, 121301(R) (2019).
\bibitem{YZhou19-a}Y. Zhou and L.W. Chen, ApJ. \textbf{886}, 52 (2019).

\bibitem{Som20}R. Somasundaram, C. Drischler, I. Tews, and J. Margueron, Phys. Rev. C {\bf 103}, 045803 (2021).

\bibitem{Mon17} C. Mondal, B.K. Agrawal, J.N. De, S.K. Samaddar, M. Centelles, and X. Vi\~{n}as, Phys. Rev. C \textbf{96}, 021302(R) (2017).

\bibitem{Bur21}C.P. Burgess, \textit{Introduction to Effective
Field Theory}, Cambridge University Press, 2021, Part I and Chap. 8.


\bibitem{JXu20}J. Xu \textit{et al.}, Phys. Lett. \textbf{B810}, 135820 (2020).
\bibitem{Zha13}Z. Zhang and L.W. Chen, Phys. Lett. \textbf{B726}, 234 (2013).
\bibitem{Cai20}B.J. Cai and B.A. Li, Phys. Rev. C {\bf 103}, 034607 (2021).

\bibitem{Tod05}B.G. Todd-Rutel and J. Piekarewicz, Phys. Rev. Lett. \textbf{95}, 122501 (2005).

\bibitem{Garg18} U. Garg and G. Col\`{o}, Prog. Part. Nucl. Phys. \textbf{101}, 55 (2018).
\bibitem{You99}D.H. Youngblood, H.L. Clark, and Y.-W. Lui, Phys. Rev. Lett. \textbf{82}, 691 (1999).
\bibitem{Shl06}S. Shlomo, V.M. Kolomietz, and G. Colo, Eur. Phys. J. A \textbf{30}, 23 (2006).
\bibitem{Che12} L.W. Chen and J.Z. Gu, J. Phys. G \textbf{39}, 035104 (2012).
\bibitem{Col14} G. Colo, U. Garg, and H. Sagawa, Eur. Phys. J. A \textbf{50}, 26 (2014).


\bibitem{Cai17x}B.J. Cai and L.W. Chen, Nucl. Sci. Tech. \textbf{28}, 185 (2017).
\bibitem{XieLi-JPG}W.J. Xie and B.A. Li,  J. Phys. G \textbf{48}, 025110 (2021).
\bibitem{Dan14} P. Danielewicz and J. Lee, Nucl. Phys. \textbf{A922}, 1 (2014).
\bibitem{XieLi-PRC21}W.J. Xie and B.A. Li, Phys. Rev. C \textbf{103}, 035802 (2021).

\bibitem{Xiao} Z. Xiao, B.-A. Li, L.-W. Chen, G.-C. Yong, and M. Zhang, Phys. Rev. Lett. {\bf 102}, 062502 (2009).

\bibitem{Feng12} Z.Q. Feng, Nucl. Phys. A {\bf 878}, 3 (2012); Phys. Lett. B {\bf 707}, 83 (2012).
\bibitem{Xie14} W.J. Xie and F.S. Zhang, Phys. Lett. B \textbf{735}, 250 (2014).

\bibitem{MSU} G. Jhang et al., Physics Letters B {\bf 813}, 136016 (2021).

\bibitem{Liu} Yangyang Liu, Yongjia Wang, Ying Cui, Chen-Jun Xia, Zhuxia Li, Yongjing Chen, Qingfeng Li, Yingxun Zhang, Phys. Rev. C {\bf 103}, 014616 (2021).

\bibitem{Kut93} M. Kutschera, W. W\'{o}jcik, Phys. Rev. C \textbf{47}, 1077 (1993).

\bibitem{Kub99} S. Kubis and M. Kutschera, AcPPB {\bf 30}, 2747 (1999).

\bibitem{Kut00} M. Kutschera and J. Niemiec, Phys. Rev. C {\bf 62}, 025802 (2000).

\bibitem{wen} D.H. Wen, B.A. Li, L.W. Chen, Phys. Rev. Lett. \textbf{103}, 211102 (2009).

\bibitem{NPnews} B.A. Li, Nuclear Physics News, {\bf 27}, 7 (2017).

\bibitem{PKU-Meng} H. Tong, P.W. Zhao, and J. Meng,  Phys. Rev. C {\bf 101}, 035802 (2020).

\bibitem{Diego} D. Lonardoni, I. Tews, S. Gandolfi, J. Carlson, Phys. Rev. Res. {\bf 2}, 022033(R) (2020).

\bibitem{Ohio20}C. Drischler, R.J. Furnstahl, J.A. Melendez and D.R. Phillips, Phys. Rev. Lett. {\bf 125}, 202702 (2020).


\end{references}
\end{document}